\documentclass[reprint,aps,prl,amsmath,amssymb,twocolumn,nobalancelastpage,citeautoscript,floatfix,raggedbottom,longbibliography]{revtex4-2}

\usepackage[usenames,dvipsnames]{xcolor}
\usepackage{graphicx,microtype}
\usepackage[bookmarks=false,colorlinks]{hyperref}
\hypersetup{hypertexnames=false,linkcolor=RubineRed,citecolor=RoyalBlue,filecolor=Mulberry,urlcolor=RoyalBlue}

\usepackage{chemformula}

\usepackage{nicefrac}
\usepackage{siunitx}\DeclareSIUnit\angstrom{\text{Å}}

\newcommand{\supf}{\textcolor{Maroon}{Figure S}}
\newcommand{\supt}{\textcolor{Maroon}{Table S}}




\newcommand{\FtoG}{\ensuremath{\textrm{F}\rightarrow\Gamma} }
\newcommand{\Fpkx}{\ensuremath{\textrm{F}+k_x} }

\begin{document}

\title{Accidental persistent spin textures in the proustite mineral family}

\author{Sean Koyama}
\email{s.koyama@u.northwestern.edu}
\affiliation{Department of Materials Science and Engineering, Northwestern University, Evanston, Illinois  60208, USA}

\author{James M.\ Rondinelli}
\email{jrondinelli@northwestern.edu}
\affiliation{Department of Materials Science and Engineering, Northwestern University, Evanston, Illinois  60208, USA}

\begin{abstract}
    Persistent spin textures (PSTs) in momentum space have the potential to enable spintronic devices which are currently limited by low spin lifetimes in nonmagnetic spin-orbit coupled materials. We perform a first-principles study on the proustite mineral family, \ch{Ag3BQ3} (B=As,\,Sb; Q=S,\,Se), and show these chalcogenides exhibit a non-symmetry protected PST, which we refer to as symmetry-assisted PSTs. Chemical substitution can be used to tune the PST quality and properties, e.g., spin lifetime, and we find that a Rashba anisotropy criterion correlates with the PST area and spin lifetime for two of the three proustites examined. Last, we show that a first-order effective SOC Hamiltonian, often used in two-dimensional systems, is insufficient to describe the PST state in all proustites, suggesting that higher order models are necessary to fully describe PSTs in bulk three-dimensional materials.
\end{abstract}

\maketitle

%\section{Introduction}

Nonmagnetic materials with unidirectional spin-momentum locking, or a  persistent spin texture (PST), exhibit electrical transport properties useful for spintronic applications \cite{bernevig_exact_2006, schliemann_colloquium_2017}. In particular, a PST at the Fermi level of a conducting system enables the persistent spin helix (PSH) transport mode in which an electron's spin precesses around a fixed axis as it moves through the material. This mode is immune to decoherence from some low-temperature scattering mechanisms. PST formation requires the presence of strong spin-orbit coupling (SOC) combined with inversion symmetry breaking. This combination produces SOC-derived band splitting and is responsible for the well-known Rashba and Dresselhaus splittings and spin textures. In contrast to the aforementioned spin textures, a PST consists of unidirectionally aligned spins which are momentum-independent. %The PST property 
This feature has been reported theoretically and observed experimentally in quantum well structures, 2D materials, and at interfaces \cite{koralek_emergence_2009, absor_highly_2021, yamaguchi_strain-induced_2017}. PSTs were recently theorized to exist in bulk three-dimensional (3D) systems exhibiting nonsymmorphic symmetries \cite{tao_persistent_2018} along with strong SOC and broken inversion symmetry, and several bulk materials, primarily ferroelectric oxides, have since been predicted to host a PST \cite{autieri_persistent_2019, djani_rationalizing_2019, lu_discovery_2020}.

The 3D PST materials predicted thus far all exhibit a symmetry-protected PST (SP-PST), also known as a Type-I PST \cite{tao_perspectives_2021,Lu/Rondinelli:2022}, which is enforced through crystalline symmetries. Recently, mirror symmetries were shown to be the key ingredient for forming a SP-PST \cite{lu_discovery_2020}. The symmetry protection, however, is a sufficient but not a necessary component to form a PST in bulk materials -- it is possible to tune the strength of spin-orbit coupling parameters  to produce an ``accidental'' or Type-II PST that is not enforced by symmetry \cite{Lu/Rondinelli:2022}. In this work, we perform electronic structure simulations and model Hamiltonian calculations to show that the proustites (\ch{Ag3BQ3}), a family of silver chalcogenide minerals, exhibit Type-II PSTs. By looking beyond symmetry-constrained PSTs and examining a family with high chemical tunability, our study expands the number of polar compounds that can exhibit high quality PSTs and spin helices.

%\section{Materials and Methods}

%\subsection{Proustite Family}

\begin{figure}
    \centering
    \includegraphics[width=0.48\textwidth]{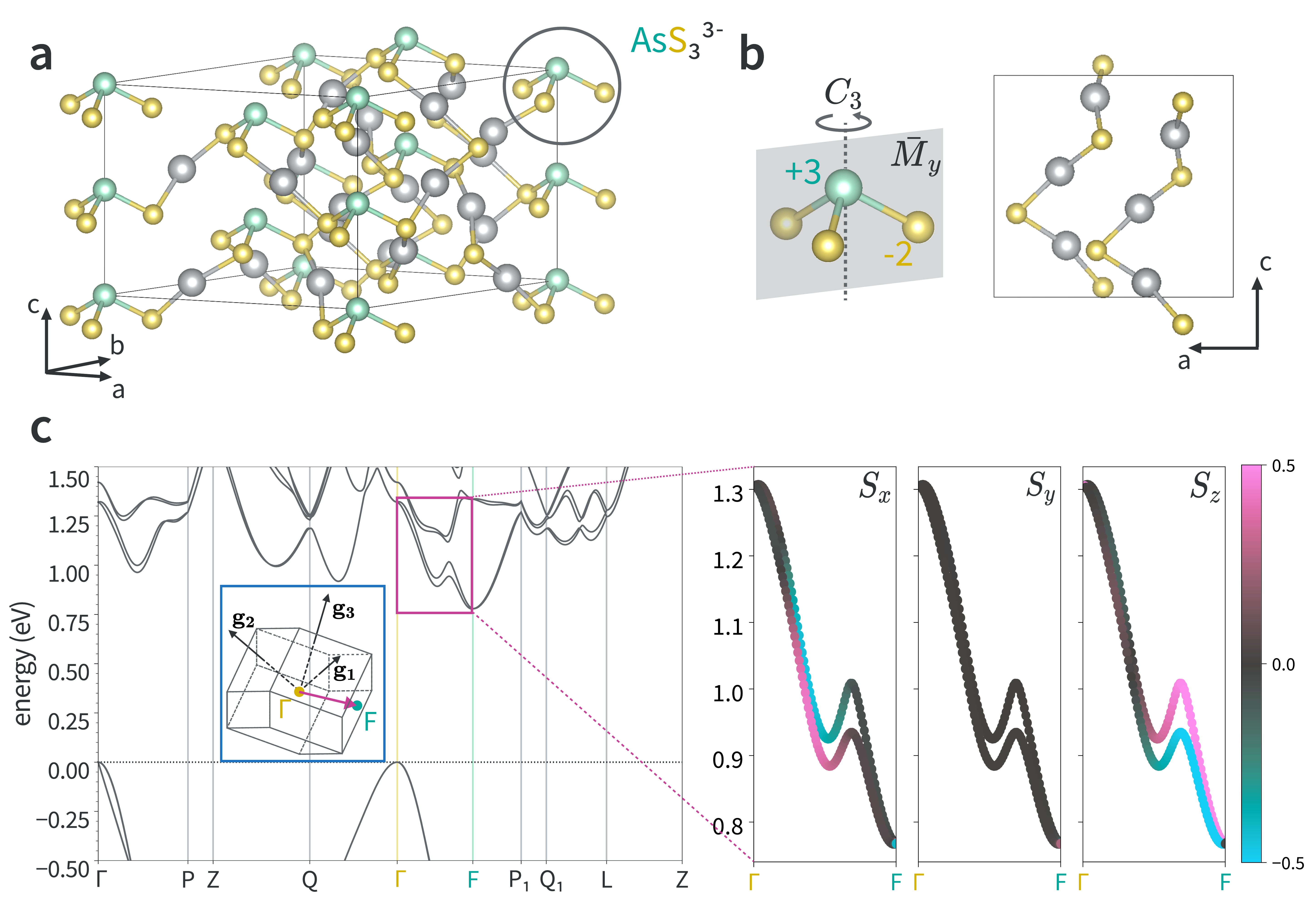}
    \caption{(a) Crystal structure of \ch{Ag3AsS3} in the hexagonal setting (b) comprises two building blocks: \ch{AsS3} pyramids linked together by \ch{Ag-S} chains. (c) The band structure of \ch{Ag3AsS3} showing the valence band minimum (VBM) and conduction band maximum (CBM) at $\Gamma$ and F, respectively. Inset: The Brillouin zone of the material with special points of interest. Zoomed: Lowest two conduction bands in the \FtoG path with projected spin character (colormapped).
    }
    \label{fig:structure_and_bands}
\end{figure}

The proustite family consists of three compounds in space group $R3c$: the namesake mineral \ch{Ag3AsS3}, as well as \ch{Ag3SbS3} and \ch{Ag3AsSe3}. The former two are naturally occurring minerals and all three have been experimentally synthesized as bulk single crystals \cite{schonau_high-temperature_2002, ewen_raman_1983, kihara_refinements_1986}. The crystal structure of \ch{Ag3AsS3} (\autoref{fig:structure_and_bands}a) consists of \ch{AsS3} pyramids connected along the $c$ axis by \ch{Ag-S} chains (\autoref{fig:structure_and_bands}b) which are generated by a $3_1$ screw axis. Owing to the $c$-glide plane, two sublattices of the \ch{AsS3} pyramids and chains exist, where each sublattice is chiral but of opposite handedness from the other, resulting in an achiral crystal. The isostructural variants are similar 
%share the same crystal structures with some differences between their lattice constants %(\autoref{tab:structure})
(see \supt1 of the Supporting Information \cite{Supp} and Ref.\ \onlinecite{structures}).

\ch{Ag3AsS3} is a semiconductor with an experimental bandgap of $E_g=\SI{1.99}{\electronvolt}$ \cite{rud_development_2010}. Its computed bandgap at the DFT-PBEsol level (see 
\footnote{Density functional theory (DFT) calculations were performed using the Vienna \textit{ab-initio} simulation package (VASP)  \cite{kresse_ab_1993, kresse_efficient_1996, kresse_efficiency_1996}  with a plane wave cutoff of \SI{350}{\electronvolt} and projector-augmented wave (PAW) pseudopotential\cite{kresse_ultrasoft_1999, blochl_projector_1994} with Ag $5s$ and $4d$; As $4s$ and $4p$; Sb $5s$ and $5p$; S $3s$ and $3p$; and Se $4s$ and $4p$ electrons as valence states. We utilized the PBEsol exchange-correlation functional with spin-orbit coupling included, unless specified otherwise  \cite{perdew_generalized_1996, perdew_restoring_2008}. The Brillouin zone is sampled with a $4\times4\times4$ $k$-point mesh and integrations performed with the tetrahedon method. 
Structures were relaxed until forces were below \SI{1e-4}{\electronvolt\per\angstrom}. Electric polarizations were calculated using the Berry phase method \cite{king-smith_theory_1993}. The HSE06 hybrid functional \cite{krukau_influence_2006} was used for accurate bandgap calculations and dense $k$-point meshes were constructed for non-self-consistent field band dispersion and spin texture calculations. The Atomic Simulation Environment (ASE) was used to aid calculations and post-processing \cite{hjorth_larsen_atomic_2017}; LOBSTER for density of states calculations \cite{dronskowski_crystal_1993, deringer_crystal_2011, maintz_analytic_2013, maintz_lobster_2016}; and VESTA \cite{momma_vesta_2011} for structure visualization.
} for computational details) 
is significantly smaller than the experimental value \supt2. % (\autoref{tab:bandgaps}). 
The inclusion of spin-orbit coupling (SOC) reduces the computed bandgap further; however, the HSE06 functional with SOC predicts a value of $E_g=\SI{1.57}{\electronvolt}$, which is closer to the experimental value. Thus, PBEsol  underestimates the bandgap by approximately 61\%, while the hybrid functional underestimates it by 21\%.
Although experimental bandgaps are not available for the other proustite variants, their computed bandgaps follow similar trends with respect to functional choice.
%and retain their relative orderings.

%\begingroup
%\squeezetable
%\begin{table}
%    \centering
%    \caption{Experimental (Exp.) and DFT bandgaps ($E_g$) of the proustite-structured compounds in units of eV. 
    %The PBEsol functional underestimates the bandgaps by approximately 61\%, while the HSE06 hybrid functional underestimates by 21\% for \ch{Ag3AsS3}.
%    }
%    \begin{ruledtabular}
%    \begin{tabular}{l c c c c}
%     & & \multicolumn{3}{c}{$E_g$ (DFT)} \\[0.2em]
%    \cline{3-5}\\[-0.8em]
%    Material & $E_g$(Exp.) & PBEsol & +SOC & +SOC+HSE06\\
%         \hline\\[-0.8em]
%        \ch{Ag3AsS3} & 1.99 \cite{rud_development_2010} & 0.782 & 0.767 & 1.57\\
%        \ch{Ag3SbS3}   & -             & 0.372 & 0.361 & 1.07\\
%        \ch{Ag3AsSe3}  & -             & 0.446 & 0.413 & 1.14\\
%    \end{tabular}
%    \end{ruledtabular}
%    \label{tab:bandgaps}
%\end{table}
%\endgroup

The DFT-computed band structure with SOC (\autoref{fig:structure_and_bands}c) reveals an indirect bandgap with its valence band maximum at $\Gamma$ and its conduction band minimum at F $(\nicefrac{1}{2}, \nicefrac{-1}{2}, 0)$. This is consistent with prior experimental reports of an indirect bandgap \cite{rud_development_2010}. The lowest conduction band in the \FtoG path shows strong spin polarization near the F point with $s_y \approx 0$ (\autoref{fig:structure_and_bands}c). This suggests that an interesting spin texture may exist near the F point; we show later that this is a PST.

\begin{figure}
    \centering
    \includegraphics[width=0.50\textwidth]{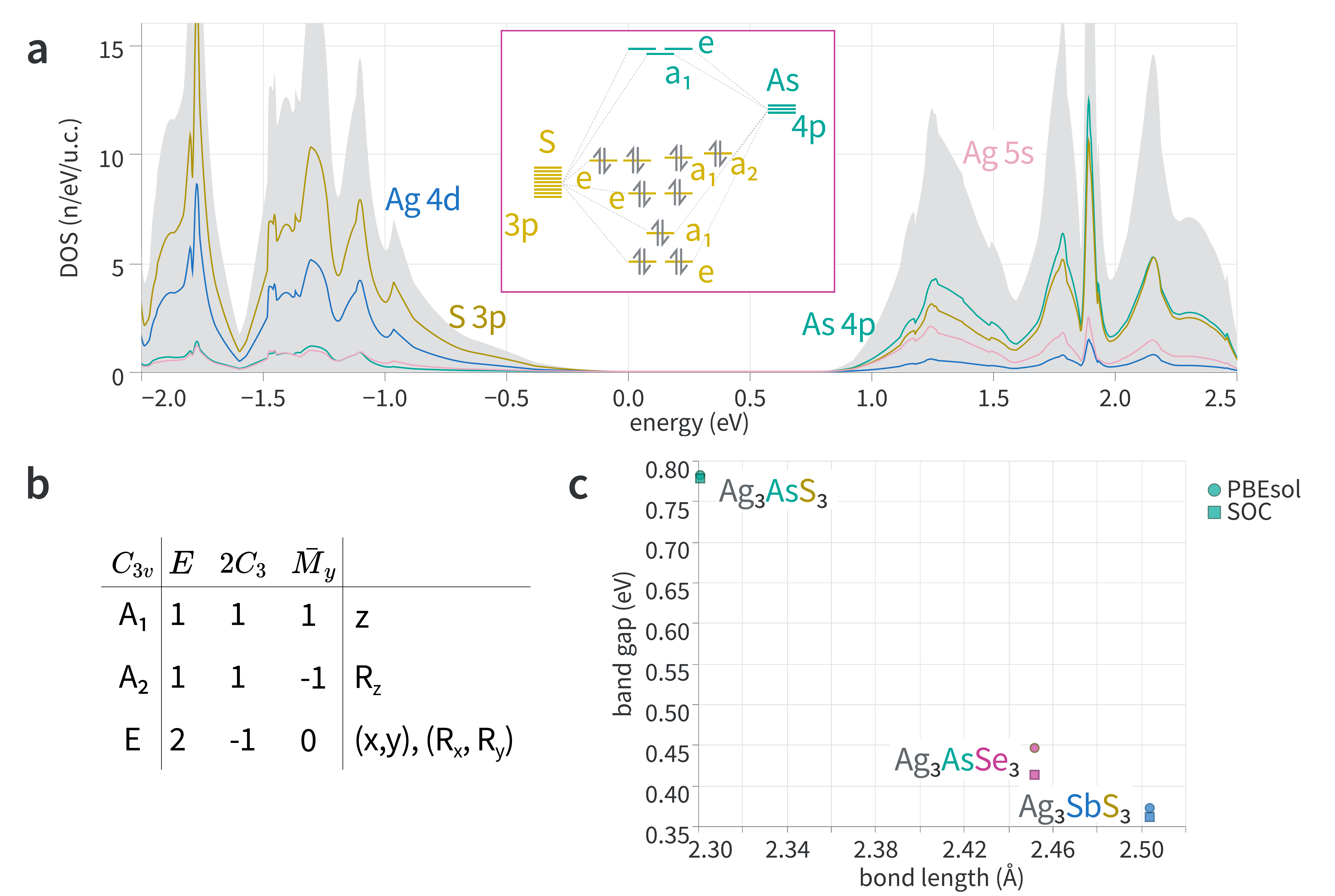}
    \caption{%
    (a) Orbital projected density of states near the Fermi level. The inset shows the molecular orbital diagram of the \ch{AsS3^{3-}} unit in \ch{Ag3AsS3} which shows the origin of the As and S orbital characters at the band edges. 
    (b) Character table of the $C_{3v}$ point group of \ch{AsS3^{3-}}.
    (c) Bandgap dependence on  As-S bond length in the proustite family showing that shorter bond lengths (stronger As-S interactions) are linearly correlated with larger bandgaps.
    }
    \label{fig:dos}
\end{figure}

The orbital-projected density-of-states  (DOS)  shows that the valence band is made up of primarily hybridized Ag $4d$ and S $3p$ orbitals (\autoref{fig:dos}a), while the conduction band consists of hybridized As $4p$ and S $3p$ with Ag $5s$ orbitals. The band edges are dominated by S and As $p$ orbitals. To elucidate the nature of the frontier orbitals forming these bands, we construct a molecular orbital (MO) diagram of the \ch{AsS3} molecular units with their $p$ orbitals as basis functions (\autoref{fig:dos}a, inset). Since the \ch{AsS3} unit has $C_{3v}$ symmetry, the $\sigma$ interactions form pairs of bonding and antibonding $a_1$ and $e$ orbitals as deduced from the character table (\autoref{fig:dos}b). The remaining S $p$ orbitals remain non-bonding with the As orbitals and form the valence band through interactions with the Ag $d$ orbitals. 
%Comparing the molecular orbitals to the band structure, we see that 
%
The MO description accurately predicts the orbital characters at the highest occupied energy level (the valence band maximum, VBM) and the lowest unoccupied energy level (the conduction band minimum, CBM).

\begin{figure*}
    \centering
    \includegraphics[width=0.75\textwidth]{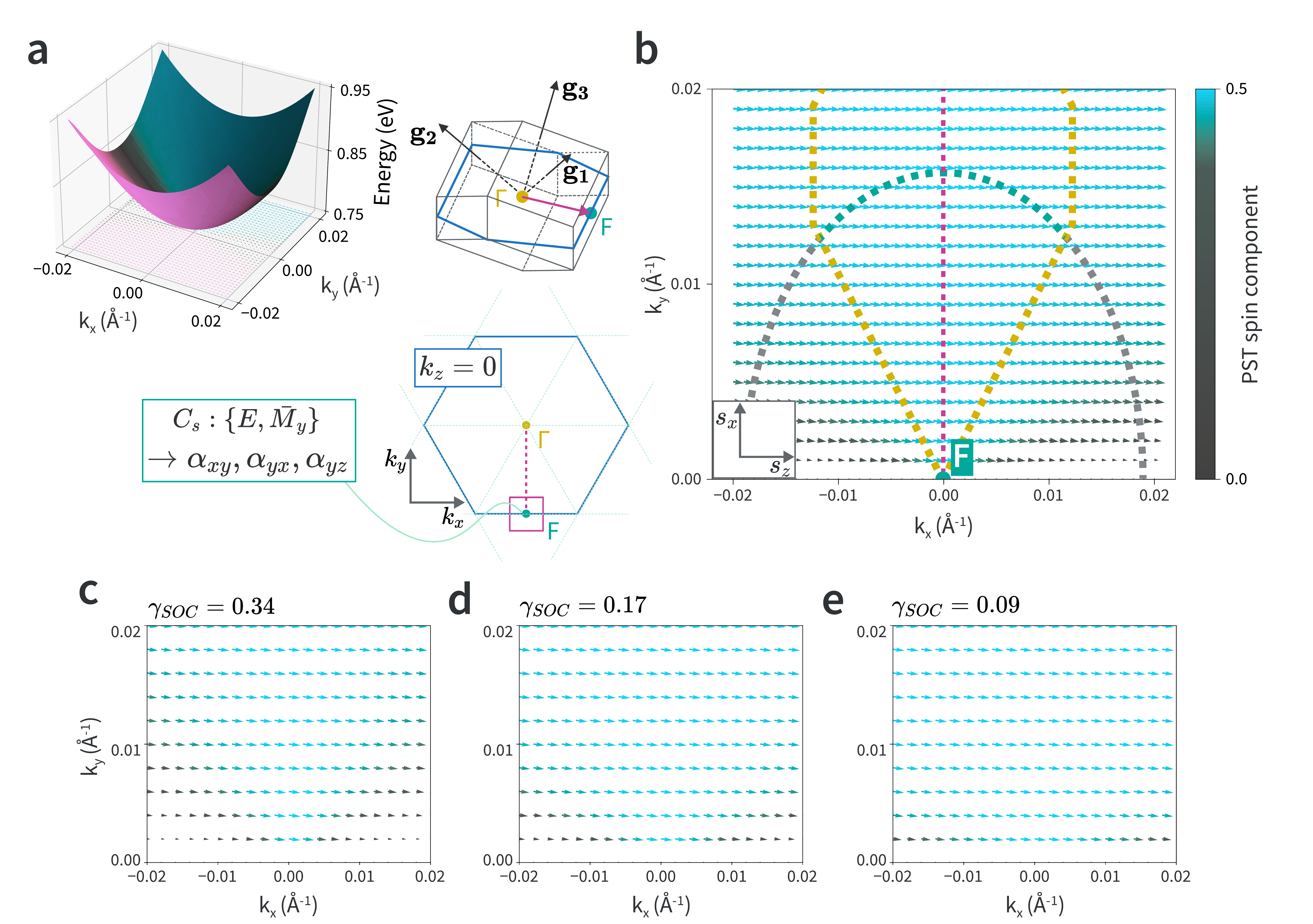}
    \caption{(a) The lowest conduction band of \ch{Ag3AsS3} in the $k_z = 0$ plane in the Brillouin zone. The full Brillouin zone and the $k_z=0$ plane are shown to the right with the \FtoG path highlighted. First-order SOC Hamiltonian terms in \autoref{eq:soc} are derived from the little group symmetries at the high-symmetry point of interest. (b) The spin texture of \ch{Ag3AsS3} showing a PST extending in the $k_y$ direction along the \FtoG path. The colormap indicates the spin-deviation angle from the PST direction. The yellow-bordered region shows where the spin deviation is $< \SI{10}{\degree}$ and the elliptical region is the Fermi arc at a doping level of $n=\SI{1e19}{\per\cubic\cm}$. Panels (c), (d), and (e) show the spin texture modeled with a two-band model (\autoref{eq:soc}) with varying values of Rashba anisotropy $\gamma_{SOC}$. Larger $\gamma_{SOC}$ leads to a smaller PST region. Panel (d) shows the model spin texture obtained using parameters fitted to the DFT data for \ch{Ag3AsS3}.}
    \label{fig:spin_textures}
\end{figure*}

The MO diagram grants us insight into the chemical origin of the band gap: the interaction between the As and S atoms. If the orbital overlap is increased (decreased) between the  $p$ orbitals of the two species, we would expect the bandgap to increase (decrease) due to further destabilization (stablization) of the anti-bonding orbitals. Since the proustite variants are isoelectronic and isostructural, the frontier orbitals are the same and we should expect the same mechanism to control the bandgap. \autoref{fig:dos}c shows this linear dependence of the band gap on the As-S (or equivalent) bond length for each of the proustite variants, verifying this mechanism of bandgap control.

%\subsection{Spin texture}

\autoref{fig:spin_textures}a shows the 
the spin character of the lowest conduction band projected onto the $k_z=0$ plane of the first Brillouin zone.
We then use this projection to  plot the spin texture of the conduction band near the F point in  \autoref{fig:spin_textures}b. 
We note that a mirror plane perpendicular to the \FtoG path intersects the F point, meaning the little group of the $k$ vector at F is $C_s$. Based on Ref.~\onlinecite{lu_discovery_2020}, we expect a PST to occur along the mirror plane ($\hat{\textrm{M}}_y$) with spin locking along $s_y$. We find, however, a PST forming near the F point with spins aligned orthogonal to $s_y$ centered along the \FtoG path and extending in the $\pm k_x$ directions. We refer to the \FtoG path as the PST path and the \emph{orthogonal} path along the mirror plane as the \Fpkx path. In addition to the unexpected direction of the PST path, we observe a small region on either side of the \Fpkx path where the spin deviates from the dominant PST direction (see color scale). The dominant PST spin orientation has normalized spin components $(0.069, 0.495)$, given as an ordered pair $(s_x, s_z)$; this oblique PST direction relative to the reciprocal lattice vectors is also unusual compared to previously reported bulk PST materials.

%\subsection{Two band model}

We now perform a symmetry analysis of the band structure of proustite near the F point to describe the SOC  characteristics and attain a better understanding of the origin of the PST. Proustite exhibits $R3c$ symmetry and the little group of the F point is $C_s$, \emph{i.e.}, the only symmetry operators are the identity and mirror plane: \{$E$, $\hat{\textrm{M}}_y$\}. Following previous symmetry analyses \cite{tao_persistent_2018, autieri_persistent_2019}, we  construct a two-band model by applying symmetry constraints to obtain a first-order SOC Hamiltonian, with the F point as the origin:
\begin{equation}
    \mathcal{H}_{SOC} = \alpha_{xy} k_x \sigma_y + \alpha_{yx} k_y \sigma_x + \alpha_{yz} k_y \sigma_z \,.
    \label{eq:soc}
\end{equation}
\autoref{eq:soc} allows us to immediately discern how the observed spin-momentum locking results along the \FtoG and \Fpkx paths. Along the \Fpkx path, $k_y=0$ and the only relevant term couples $k_x$ with $\sigma_y$, which explains the locking of the spin direction along the $s_y$ axis on this path. Along the \FtoG path, $k_x=0$ and the relevant terms couple $k_y$ to both $\sigma_x$ and $\sigma_z$, explaining the oblique spin texture along this path. 

We assume that the full Hamiltonian describing the system near the conduction band minimum takes the form $\mathcal{H} = \mathcal{H}_0 + \mathcal{H}_{SOC}$, with $\mathcal{H}_{SOC}$ given by \autoref{eq:soc} and $\mathcal{H}_0$ being the free electron Hamiltonian,
$    \mathcal{H}_0 = \hbar^2 k^2 / 2m^*  $.
Although $ \mathcal{H}_0$ is a poor approximation of the real system as a whole, the approximation holds utility as an analytic tool to understand the mechanisms of the PST in proustite since the CBM can be locally approximated as parabolic. Since the Hamiltonian is quadratic in $k$ and is otherwise only dependent on system-specific parameters, the Schr\"odinger equation can be solved analytically to give the following eigenvalues:

\begin{equation}
    E_{\pm}(k) = \mathcal{H}_0 \pm \sqrt{(\alpha_{xy} k_x)^2 + (\alpha_{yx} k_y)^2 + (\alpha_{yz} k_y)^2}\,.
\end{equation}

\begin{table*}
    \centering
    \caption{Polarization ($P$), SOC parameters ($\alpha_{ij}, \gamma_{SOC}$), and PSH properties (length $l_{PSH}$ and period $ T_{PSH}$) for proustite and its variants. The Rashba anisotropy $\gamma_{SOC}$ is the ratio of the coupling strength along $x$ to that along $y$. A smaller (more pronounced) Rashba anisotropy is favorable to PST formation and larger spin lifetime. {\ch{Ag3AsSe3}} has no computed spin lifetime ($\tau_s$), indicated by N/A, because its conduction band minimum lies outside of the PST region, making it practically inaccessible.}
    \begin{ruledtabular}
    \begin{tabular}{l l l l l l l l l}
        Material    & $P$ ($\si{\micro\coulomb}/\si{\centi\,meter^{-2}}$) & $\alpha_{xy} (\si{eV\angstrom})$ & $\alpha_{yx}$ (\si{eV\angstrom}) & $\alpha_{yz}$ (\si{eV\angstrom}) & $\gamma_{SOC}$ & $l_{PSH}$ (\si{nm}) & $\tau_s$ (\si{ps}) & $\tau_s / T_{PSH}$\\
        \hline
        \ch{Ag3AsS3}     & 71.9 & 0.195 & 0.155 & -1.12 & 0.173 & 231 & 2.00 & 3.18 \\
        \ch{Ag3AsSe3}    & 109  & 0.210 & 1.192 & 1.81  & 0.0972& N/A & N/A & N/A \\
        \ch{Ag3SbS3}     & 65.1 & 0.166 & 0.539 & -1.92 & 0.0833& 148 & 3.69 & 21.7 \\
    \end{tabular}
    \end{ruledtabular}
    \label{tab:soc}
\end{table*}

Along each \FtoG and \Fpkx path, one of the momentum components $k_i$ is zero and the energy dispersion takes the form of a simple parabola with a linear coupling term. We can also solve for the eigenstates and their corresponding spin expectation values; by doing so, we are able to fit the computed DFT data to extract $\alpha_{ij}$ coupling coefficients and plot the resulting spin textures to assess whether the model reproduces the DFT-simulated PST.
\autoref{tab:soc} presents the $\alpha_{ij}$ values obtained by fitting to the DFT data about the F point. We define an effective $\alpha_{y,\mathit{eff}}= (\alpha_{yx}^2 + \alpha_{yz}^2)^{1/2}$ to describe the net SOC band shift along $k_y$ and note that there is an order of magnitude difference between the SOC coupling strengths along $k_x$ and along $k_y$. This variation  explains the dominance of the spin texture along $k_y$. Because the SOC is much larger along the \FtoG path, the associated spin texture dominates the larger 2D region of the Brillouin zone.

We show the spin texture resulting from the two band model with the fitted parameters in \autoref{fig:spin_textures}d. We see that the first-order approximation fits the computed spin texture well in the region close to the F point. In \autoref{fig:spin_textures}c-e, we vary the SOC parameters in the two band model to show how the parameters affect the PST. We find that the PST is controlled by the ratio $\gamma_{SOC} = \alpha_{xy} / \alpha_{y,\mathit{eff}}$; this ratio is equivalent to the Rashba anisotropy identified in Ref.~\onlinecite{lu_discovery_2020} as a predictor of PST quality in Type-I PSTs. A high degree of Rashba anisotropy (small $\gamma_{SOC}$) leads to the suppression of the weakly coupled spin component which results in a large PST area. As $\gamma_{SOC} \rightarrow 0$, we approach a perfect PST. $\gamma_{SOC}=1$ corresponds to a case with no anisotropy between the SOC parameters, and the model reproduces a Rashba or Dresselhaus type spin texture, which has no inherent PST character. Thus, we have identified and further confirmed the  first-order Rashba anisotropy as one of the key indicators of PST quality for symmetry-protected PSTs.

It is of particular interest that the  Rashba anisotropy is an effective predictor of PST area in both Type-I symmetry-protected PSTs, such as in Ref.~\onlinecite{lu_discovery_2020}, and Type-II accidental PSTs, as in this work. This finding blurs the lines between the current field of Type-I and Type-II PSTs, especially because if $\alpha_{xy}$ were to dominate the SOC Hamiltonian in \autoref{eq:soc}, the resulting PST would be along a mirror plane and thus potentially be categorized as a Type-I SP-PST. Consequently, the particular symmetries do not appear to dictate the quality of the PST, and instead the chemical interactions of the system and their interplay with the momentum-space physics determines the PST quality. This has been noted for some  quasi-2D layered perovskite PSTs where the structural characteristics combine with the crystalline symmetries to constrain the relevant SOC terms to one in-plane direction, producing a PST which is occasionally labeled as symmetry-protected. However, this type of symmetry-protection is distinct from that of Ref.~\onlinecite{tao_persistent_2018}, in which symmetry constraints alone guarantee the existence of a PST near certain high-symmetry points. This strict symmetry requirement for defining SP-PSTs has been disputed \cite{lu_discovery_2020} and is evidently too stringent for searching for new PST materials, as a multitude of materials without the strict symmetry protection have been shown to host a PST. Thus we propose a distinction between symmetry-\textit{protected} PSTs in which symmetry guarantees a PST to first order (e.g.,  \ch{BiInO3}\cite{tao_persistent_2018}) and symmetry-\textit{assisted} PSTs in which symmetry reduces the number of first-order SOC terms such that structural or chemical features may force the system into a PST (e.g.,  layered perovskites \cite{lu_discovery_2020, zhang_room-temperature_2022}). The proustite PSTs presented here  fit into the latter category, although there is a significant distinction in the PST formation mechanism in quasi-2D perovskites (see Ref. \onlinecite{zhang_room-temperature_2022}) and the materials presented here. The layered perovskites exhibit PSTs due to structural contraints -- since the polarization is in-plane and the structure is quasi-2D, spin splitting occurs largely along one in-plane direction, producing a PST. In the 3D  proustites, there is no intrinsic structural constraint. The large  Rashba anisotropy forms exclusively through orbital interactions. Both cases, however, are notably distinct from true symmetry-protected PSTs by the fact that symmetry only plays a partial role in producing a symmetry-assisted PST.

Additionally, the result of our model is in contrast to the analysis presented in Ref.~\onlinecite{autieri_persistent_2019} for the layered perovskite \ch{CsBiNb2O7}, where the \emph{cubic} SOC splitting terms dominate and the anisotropy between those terms determines the PST quality. The difference in results can be explained by the difference in band dispersions. Since the PST in proustite is observed near the F point, the first order expansion is sufficient to describe the PST near it. In \ch{CsBiNb2O7}, the band extremum and associated PST is seen at large momentum, where higher order terms dominate and are thus necessary as part of the analysis. 

\begin{figure}
    \centering
    \includegraphics[width=0.5\textwidth]{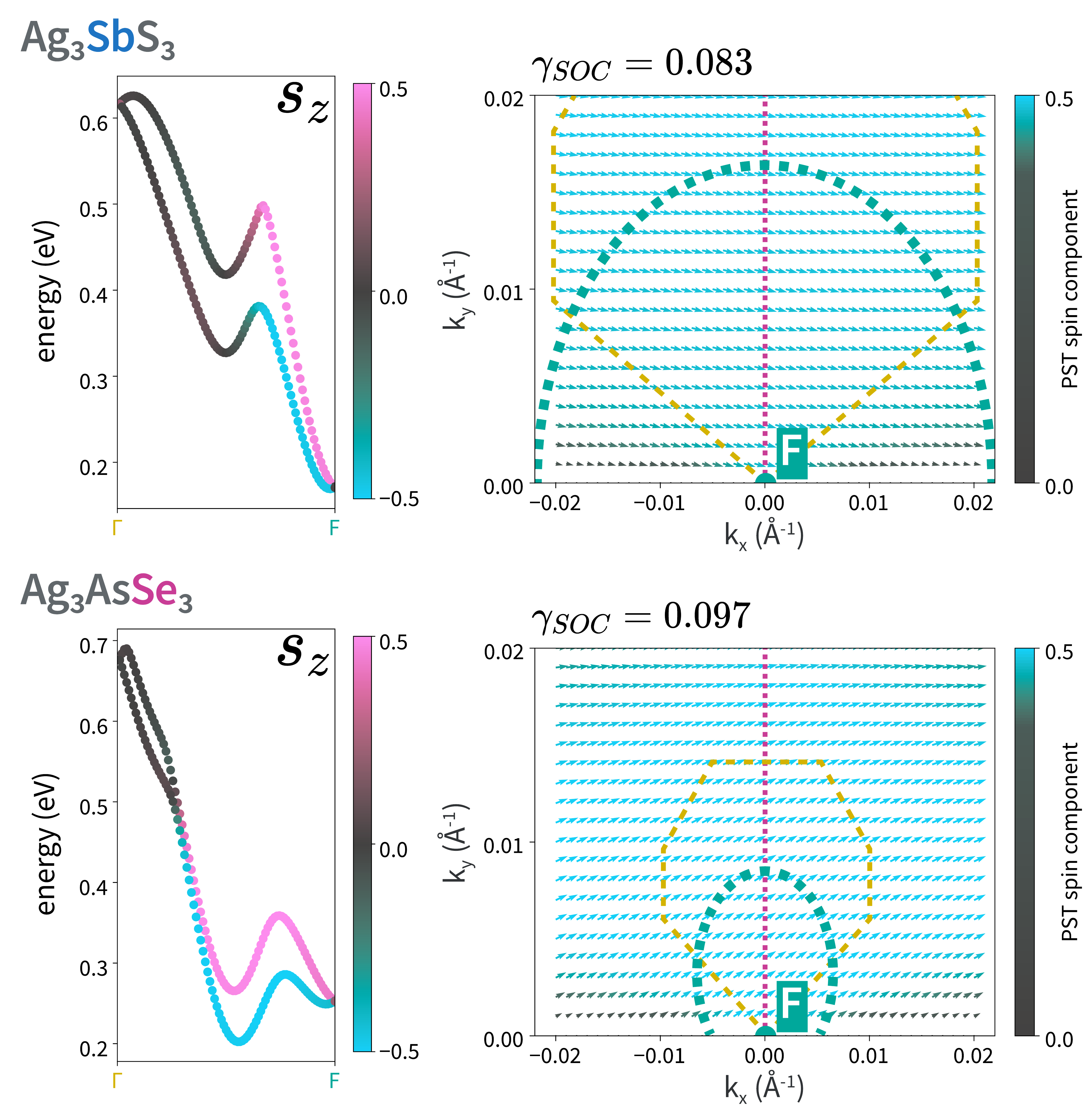}
    \caption{The $s_z$-projected conduction band and spin textures of \ch{Ag3SbS3} and \ch{Ag3AsSe3} near the F point. Both variants display PSTs with \ch{Ag3SbS3} possessing a significantly greater PST area compared to \ch{Ag3AsS3}. In the case of \ch{Ag3AsSe3}, the CBM shifts due to bands shifting to lower energy along both the \FtoG path and the $\Gamma\rightarrow\textrm{M} (1/3, -2/3, 1/3)$ paths (\supf2); these two band minima are nearly degenerate within $\sim$\SI{10}{\milli\electronvolt} of each other. Consequently, the PST is inaccessible in \ch{Ag3AsSe3}.}
    \label{fig:variants}
\end{figure}

%\subsection{Proustite variants}
We now examine the two isostructural and isoelectronic proustite variants: \ch{Ag3SbS3} and \ch{Ag3AsSe3}. 
%These variants share the same structure and have similar bonding characteristics as proustite. Owing to the isostructural and isoelectronic nature of the variants, 
Although they are chemically similar, there are  significant changes in the spin-orbit coupling strengths. In addition, the location of the CBM changes drastically in \ch{Ag3AsSe3}.
\autoref{fig:variants} shows 
%We show 
the SOC band dispersions and PSTs of the conduction band of these variants; %in \autoref{fig:variants}; 
the corresponding SOC parameters are tabulated in \autoref{tab:soc}. The Rashba anisotropy is greatest in \ch{Ag3SbS3} followed by \ch{Ag3AsSe3}. The PST areas outlined in yellow indicate regions within which the spin direction does not deviate from the PST beyond \SI{10}{\degree}; the larger the area with small spin deviation, the closer the PST is to ideal. We see that immediate to the F point ($k_y < \SI{0.01}{\per\angstrom}$), $\gamma_{SOC}$ is a good predictor of PST area -- smaller $\gamma_{SOC}$ leads to a larger area. In \ch{Ag3AsSe3}, $\gamma_{SOC}$ fails as a predictor of PST area, although the  fraction of the Fermi arc in the PST region is still large. This failure of $\gamma_{SOC}$ in predicting PST area is likely due to large cubic contributions to the SOC Hamiltonian as in \ch{CsBiNb2O7} \cite{autieri_persistent_2019}. This again suggests that first-order terms are insufficient to fully describe the PST in all materials, particularly those where the band minimum lies far from a high-symmetry point.

An important note is the location of the conduction band minimum. The PST states must lie at the CBM since they are transport states which need to be electron doped to access. While the band structure shown suggests that the conduction band minimum of \ch{Ag3AsSe3} lies within the \FtoG path, it is rather located in the $\Gamma\rightarrow\textrm{M} (\nicefrac{1}{3}, \nicefrac{-2}{3}, \nicefrac{1}{3})$ path (\supf2 of Ref.\ \onlinecite{Supp}). The CBM seen in the \FtoG path is nearly degenerate with the true CBM. This path lies outside of the useful PST region, making the PST in \ch{Ag3AsSe3} practically inaccessible.

The PSTs of the proustite variants confirm that the first-order Rashba anisotropy is a predictor of PST quality in materials where the CBM lies near the high symmetry point. We further confirm this by computing the spin lifetime $\tau_s$ of the resulting persistent spin helix and the ratio of spin lifetime to the PSH period, which is a limiting factor in the material's device application potential (see Refs.\ \onlinecite{Supp} and \onlinecite{liu_unified_2012} for details). We see that in the proustite family, \ch{Ag3SbS3} has the longest spin lifetime (\SI{3.69}{ps}) and $\tau_s / T_{PSH}$ ratio (\num{21.7}). However, compared to recently predicted bulk PST materials, these quantities are 1-3 orders of magnitude smaller and may require further engineering to find  practical use \cite{lu_discovery_2020}. For \ch{Ag3AsSe3}, the location of the band minimum prevents the two-band model from fully describing the PST and therefore we do not compute the spin lifetime using the first-order model.

We also recognize the necessity of finding appropriate dopants to access the PST states. The PST states lie in the conduction band so $n$-type doping is needed to access the PSH transport mode. One possible route to doping could be through vacancies on the chalcogenide site as in \ch{MoS2} \cite{shen_healing_2022}. Photodoping could also be used to probe the PST states by utilizing circularly polarized light to excite electrons into the spin-polarized bands \cite{zhang_room-temperature_2022}.
%offers potential to directly observe the PST states.

%\section{Conclusion}
%\paragraph{Conclusion.}%
Although the recent research into 3D bulk PST compounds has focused on PSTs formed by specific crystalline symmetries, such as the $C_{2v}$ point group, we showed that such symmetries are unnecessary constraints in the search for new PST materials. We showed that the proustite mineral family offers a flexible platform to design structural and electronic properties, including tuning the low-symmetry PST in the conduction band. We also distinguished symmetry-\emph{assisted} PSTs from symmetry-\emph{protected} PSTs, categorizing the majority of previously identified PST materials along with the proustites as symmetry-assisted PSTs. We propose that the Rashba anisotropy $\gamma_{SOC}$ is a useful parameter in evaluating PSTs near a high-symmetry $k$ vector and that this concept may be extended to higher orders.
%; a high degree of third-order Rashba anisotropy was previously shown to be key to producing a useful PST at high momentum $k$-vectors in a layered perovskite. 
We conjecture that high Rashba anisotropy at multiple orders could result in an ideal PST that spans a significant portion of the Brillouin zone. We propose that this Rashba anisotropy criterion could be used in high-throughput screening of materials for future bulk PST identification in a manner similar to searchers for novel Rashba materials \cite{Mera_Acosta_2020}. Since symmetry does not present a strong constraint on the existence of bulk PSTs, there may be several unexplored material classes in which PSTs are hidden. In addition, future work into studying the fundamental mechanisms of PSTs should focus on the chemical and structural origins of the Rashba anisotropy, as this is still unclear. Understanding how to control this Rashba anisotropy may be the key to unlock full design control of PST materials.

\begin{acknowledgments}
S.K.\ thanks Dr.\ X.-Z.\ Lu for his guidance and many stimulating discussions.
This research was supported by the National Science Foundation (NSF) under Award DMR-2104397. Computational resources were provided by: \texttt{Quest} high performance computing facility at Northwestern University which is jointly supported by the Office of the Provost, the Office for Research, and Northwestern University Information Technology; \texttt{Stampede2} at the Extreme Science and Engineering Discovery Environment (XSEDE), funded by the NSF through award ACI-1540931; and \texttt{Carbon} at the Center for Nanoscale Materials, a U.S. Department of Energy Office of Science User Facility, supported by the U.S. DOE, Office of Basic Energy Sciences, under Contract No. DE-AC02-06CH11357.
\end{acknowledgments}

\bibliography{reference}

\begin{thebibliography}{37}%
\makeatletter
\providecommand \@ifxundefined [1]{%
 \@ifx{#1\undefined}
}%
\providecommand \@ifnum [1]{%
 \ifnum #1\expandafter \@firstoftwo
 \else \expandafter \@secondoftwo
 \fi
}%
\providecommand \@ifx [1]{%
 \ifx #1\expandafter \@firstoftwo
 \else \expandafter \@secondoftwo
 \fi
}%
\providecommand \natexlab [1]{#1}%
\providecommand \enquote  [1]{``#1''}%
\providecommand \bibnamefont  [1]{#1}%
\providecommand \bibfnamefont [1]{#1}%
\providecommand \citenamefont [1]{#1}%
\providecommand \href@noop [0]{\@secondoftwo}%
\providecommand \href [0]{\begingroup \@sanitize@url \@href}%
\providecommand \@href[1]{\@@startlink{#1}\@@href}%
\providecommand \@@href[1]{\endgroup#1\@@endlink}%
\providecommand \@sanitize@url [0]{\catcode `\\12\catcode `\$12\catcode
  `\&12\catcode `\#12\catcode `\^12\catcode `\_12\catcode `\%12\relax}%
\providecommand \@@startlink[1]{}%
\providecommand \@@endlink[0]{}%
\providecommand \url  [0]{\begingroup\@sanitize@url \@url }%
\providecommand \@url [1]{\endgroup\@href {#1}{\urlprefix }}%
\providecommand \urlprefix  [0]{URL }%
\providecommand \Eprint [0]{\href }%
\providecommand \doibase [0]{https://doi.org/}%
\providecommand \selectlanguage [0]{\@gobble}%
\providecommand \bibinfo  [0]{\@secondoftwo}%
\providecommand \bibfield  [0]{\@secondoftwo}%
\providecommand \translation [1]{[#1]}%
\providecommand \BibitemOpen [0]{}%
\providecommand \bibitemStop [0]{}%
\providecommand \bibitemNoStop [0]{.\EOS\space}%
\providecommand \EOS [0]{\spacefactor3000\relax}%
\providecommand \BibitemShut  [1]{\csname bibitem#1\endcsname}%
\let\auto@bib@innerbib\@empty
%</preamble>
\bibitem [{\citenamefont {Bernevig}\ \emph {et~al.}(2006)\citenamefont
  {Bernevig}, \citenamefont {Orenstein},\ and\ \citenamefont
  {Zhang}}]{bernevig_exact_2006}%
  \BibitemOpen
  \bibfield  {author} {\bibinfo {author} {\bibfnamefont {B.~A.}\ \bibnamefont
  {Bernevig}}, \bibinfo {author} {\bibfnamefont {J.}~\bibnamefont
  {Orenstein}},\ and\ \bibinfo {author} {\bibfnamefont {S.-C.}\ \bibnamefont
  {Zhang}},\ }\bibfield  {title} {\bibinfo {title} {Exact {SU}(2) symmetry and
  persistent spin helix in a spin-orbit coupled system},\ }\href
  {https://doi.org/10.1103/PhysRevLett.97.236601} {\bibfield  {journal}
  {\bibinfo  {journal} {Phys. Rev. Lett.}\ }\textbf {\bibinfo {volume} {97}},\
  \bibinfo {pages} {236601} (\bibinfo {year} {2006})}\BibitemShut {NoStop}%
\bibitem [{\citenamefont {Schliemann}(2017)}]{schliemann_colloquium_2017}%
  \BibitemOpen
  \bibfield  {author} {\bibinfo {author} {\bibfnamefont {J.}~\bibnamefont
  {Schliemann}},\ }\bibfield  {title} {\bibinfo {title} {Colloquium: Persistent
  spin textures in semiconductor nanostructures},\ }\href
  {https://doi.org/10.1103/RevModPhys.89.011001} {\bibfield  {journal}
  {\bibinfo  {journal} {Rev. Mod. Phys.}\ }\textbf {\bibinfo {volume} {89}},\
  \bibinfo {pages} {011001} (\bibinfo {year} {2017})}\BibitemShut {NoStop}%
\bibitem [{\citenamefont {Koralek}\ \emph {et~al.}(2009)\citenamefont
  {Koralek}, \citenamefont {Weber}, \citenamefont {Orenstein}, \citenamefont
  {Bernevig}, \citenamefont {Zhang}, \citenamefont {Mack},\ and\ \citenamefont
  {Awschalom}}]{koralek_emergence_2009}%
  \BibitemOpen
  \bibfield  {author} {\bibinfo {author} {\bibfnamefont {J.~D.}\ \bibnamefont
  {Koralek}}, \bibinfo {author} {\bibfnamefont {C.~P.}\ \bibnamefont {Weber}},
  \bibinfo {author} {\bibfnamefont {J.}~\bibnamefont {Orenstein}}, \bibinfo
  {author} {\bibfnamefont {B.~A.}\ \bibnamefont {Bernevig}}, \bibinfo {author}
  {\bibfnamefont {S.-C.}\ \bibnamefont {Zhang}}, \bibinfo {author}
  {\bibfnamefont {S.}~\bibnamefont {Mack}},\ and\ \bibinfo {author}
  {\bibfnamefont {D.~D.}\ \bibnamefont {Awschalom}},\ }\bibfield  {title}
  {\bibinfo {title} {Emergence of the persistent spin helix in semiconductor
  quantum wells},\ }\href {https://doi.org/10.1038/nature07871} {\bibfield
  {journal} {\bibinfo  {journal} {Nature}\ }\textbf {\bibinfo {volume} {458}},\
  \bibinfo {pages} {610} (\bibinfo {year} {2009})}\BibitemShut {NoStop}%
\bibitem [{\citenamefont {Absor}\ \emph {et~al.}(2021)\citenamefont {Absor},
  \citenamefont {Faishal}, \citenamefont {Anshory}, \citenamefont {Santoso},\
  and\ \citenamefont {Ishii}}]{absor_highly_2021}%
  \BibitemOpen
  \bibfield  {author} {\bibinfo {author} {\bibfnamefont {M.~A.~U.}\
  \bibnamefont {Absor}}, \bibinfo {author} {\bibfnamefont {Y.}~\bibnamefont
  {Faishal}}, \bibinfo {author} {\bibfnamefont {M.}~\bibnamefont {Anshory}},
  \bibinfo {author} {\bibfnamefont {I.}~\bibnamefont {Santoso}},\ and\ \bibinfo
  {author} {\bibfnamefont {F.}~\bibnamefont {Ishii}},\ }\bibfield  {title}
  {\bibinfo {title} {Highly persistent spin textures with giant tunable spin
  splitting in the two-dimensional germanium monochalcogenides},\ }\href
  {https://doi.org/10.1088/1361-648x/ac0383} {\bibfield  {journal} {\bibinfo
  {journal} {Journal of Physics: Condensed Matter}\ }\textbf {\bibinfo {volume}
  {33}},\ \bibinfo {pages} {305501} (\bibinfo {year} {2021})}\BibitemShut
  {NoStop}%
\bibitem [{\citenamefont {Yamaguchi}\ and\ \citenamefont
  {Ishii}(2017)}]{yamaguchi_strain-induced_2017}%
  \BibitemOpen
  \bibfield  {author} {\bibinfo {author} {\bibfnamefont {N.}~\bibnamefont
  {Yamaguchi}}\ and\ \bibinfo {author} {\bibfnamefont {F.}~\bibnamefont
  {Ishii}},\ }\bibfield  {title} {\bibinfo {title} {{Strain-induced large spin
  splitting and persistent spin helix at LaAlO$_3$/SrTiO$_3$ interface}},\
  }\href {https://doi.org/10.7567/APEX.10.123003} {\bibfield  {journal}
  {\bibinfo  {journal} {Appl. Phys. Express}\ }\textbf {\bibinfo {volume}
  {10}},\ \bibinfo {pages} {123003} (\bibinfo {year} {2017})}\BibitemShut
  {NoStop}%
\bibitem [{\citenamefont {Tao}\ and\ \citenamefont
  {Tsymbal}(2018)}]{tao_persistent_2018}%
  \BibitemOpen
  \bibfield  {author} {\bibinfo {author} {\bibfnamefont {L.~L.}\ \bibnamefont
  {Tao}}\ and\ \bibinfo {author} {\bibfnamefont {E.~Y.}\ \bibnamefont
  {Tsymbal}},\ }\bibfield  {title} {\bibinfo {title} {Persistent spin texture
  enforced by symmetry},\ }\href {https://doi.org/10.1038/s414>67-018-05137-0}
  {\bibfield  {journal} {\bibinfo  {journal} {Nat. Commun.}\ }\textbf {\bibinfo
  {volume} {9}},\ \bibinfo {pages} {2763} (\bibinfo {year} {2018})}\BibitemShut
  {NoStop}%
\bibitem [{\citenamefont {Autieri}\ \emph {et~al.}(2019)\citenamefont
  {Autieri}, \citenamefont {Barone}, \citenamefont {Sławińska},\ and\
  \citenamefont {Picozzi}}]{autieri_persistent_2019}%
  \BibitemOpen
  \bibfield  {author} {\bibinfo {author} {\bibfnamefont {C.}~\bibnamefont
  {Autieri}}, \bibinfo {author} {\bibfnamefont {P.}~\bibnamefont {Barone}},
  \bibinfo {author} {\bibfnamefont {J.}~\bibnamefont {Sławińska}},\ and\
  \bibinfo {author} {\bibfnamefont {S.}~\bibnamefont {Picozzi}},\ }\bibfield
  {title} {\bibinfo {title} {{Persistent spin helix in Rashba-Dresselhaus
  ferroelectric CsBiNb$_2$O$_7$}},\ }\href
  {https://doi.org/10.1103/PhysRevMaterials.3.084416} {\bibfield  {journal}
  {\bibinfo  {journal} {Phys. Rev. Materials}\ }\textbf {\bibinfo {volume}
  {3}},\ \bibinfo {pages} {084416} (\bibinfo {year} {2019})}\BibitemShut
  {NoStop}%
\bibitem [{\citenamefont {Djani}\ \emph {et~al.}(2019)\citenamefont {Djani},
  \citenamefont {Garcia-Castro}, \citenamefont {Tong}, \citenamefont {Barone},
  \citenamefont {Bousquet}, \citenamefont {Picozzi},\ and\ \citenamefont
  {Ghosez}}]{djani_rationalizing_2019}%
  \BibitemOpen
  \bibfield  {author} {\bibinfo {author} {\bibfnamefont {H.}~\bibnamefont
  {Djani}}, \bibinfo {author} {\bibfnamefont {A.~C.}\ \bibnamefont
  {Garcia-Castro}}, \bibinfo {author} {\bibfnamefont {W.-Y.}\ \bibnamefont
  {Tong}}, \bibinfo {author} {\bibfnamefont {P.}~\bibnamefont {Barone}},
  \bibinfo {author} {\bibfnamefont {E.}~\bibnamefont {Bousquet}}, \bibinfo
  {author} {\bibfnamefont {S.}~\bibnamefont {Picozzi}},\ and\ \bibinfo {author}
  {\bibfnamefont {P.}~\bibnamefont {Ghosez}},\ }\bibfield  {title} {\bibinfo
  {title} {{Rationalizing and engineering Rashba spin-splitting in
  ferroelectric oxides}},\ }\href {https://doi.org/10.1038/s41535-019-0190-z}
  {\bibfield  {journal} {\bibinfo  {journal} {npj Quantum Mater.}\ }\textbf
  {\bibinfo {volume} {4}},\ \bibinfo {pages} {51} (\bibinfo {year}
  {2019})}\BibitemShut {NoStop}%
\bibitem [{\citenamefont {Lu}\ and\ \citenamefont
  {Rondinelli}(2020)}]{lu_discovery_2020}%
  \BibitemOpen
  \bibfield  {author} {\bibinfo {author} {\bibfnamefont {X.-Z.}\ \bibnamefont
  {Lu}}\ and\ \bibinfo {author} {\bibfnamefont {J.~M.}\ \bibnamefont
  {Rondinelli}},\ }\bibfield  {title} {\bibinfo {title} {Discovery principles
  and materials for symmetry-protected persistent spin textures with long spin
  lifetimes},\ }\href {https://doi.org/10.1016/j.matt.2020.08.028} {\bibfield
  {journal} {\bibinfo  {journal} {Matter}\ }\textbf {\bibinfo {volume} {3}},\
  \bibinfo {pages} {1211} (\bibinfo {year} {2020})}\BibitemShut {NoStop}%
\bibitem [{\citenamefont {Tao}\ and\ \citenamefont
  {Tsymbal}(2021)}]{tao_perspectives_2021}%
  \BibitemOpen
  \bibfield  {author} {\bibinfo {author} {\bibfnamefont {L.~L.}\ \bibnamefont
  {Tao}}\ and\ \bibinfo {author} {\bibfnamefont {E.~Y.}\ \bibnamefont
  {Tsymbal}},\ }\bibfield  {title} {\bibinfo {title} {Perspectives of
  spin-textured ferroelectrics},\ }\href
  {https://doi.org/10.1088/1361-6463/abcc25} {\bibfield  {journal} {\bibinfo
  {journal} {Journal of Physics D: Applied Physics}\ }\textbf {\bibinfo
  {volume} {54}},\ \bibinfo {pages} {113001} (\bibinfo {year}
  {2021})}\BibitemShut {NoStop}%
\bibitem [{\citenamefont {Lu}\ and\ \citenamefont
  {Rondinelli}(2022)}]{Lu/Rondinelli:2022}%
  \BibitemOpen
  \bibfield  {author} {\bibinfo {author} {\bibfnamefont {X.-Z.}\ \bibnamefont
  {Lu}}\ and\ \bibinfo {author} {\bibfnamefont {J.~M.}\ \bibnamefont
  {Rondinelli}},\ }\href {https://doi.org/10.48550/ARXIV.2208.13862} {\bibinfo
  {title} {Strain engineering a persistent spin helix with infinite spin
  lifetime}} (\bibinfo {year} {2022})\BibitemShut {NoStop}%
\bibitem [{\citenamefont {Schönau}\ and\ \citenamefont
  {Redfern}(2002)}]{schonau_high-temperature_2002}%
  \BibitemOpen
  \bibfield  {author} {\bibinfo {author} {\bibfnamefont {K.~A.}\ \bibnamefont
  {Schönau}}\ and\ \bibinfo {author} {\bibfnamefont {S.~A.~T.}\ \bibnamefont
  {Redfern}},\ }\bibfield  {title} {\bibinfo {title} {{High-temperature phase
  transitions, dielectric relaxation, and ionic mobility of proustite,
  Ag$_3$AsS$_3$, and pyrargyrite, Ag$_3$SbS$_3$}},\ }\href
  {https://doi.org/10.1063/1.1520720} {\bibfield  {journal} {\bibinfo
  {journal} {Journal of Applied Physics}\ }\textbf {\bibinfo {volume} {92}},\
  \bibinfo {pages} {7415} (\bibinfo {year} {2002})}\BibitemShut {NoStop}%
\bibitem [{\citenamefont {Ewen}\ \emph {et~al.}(1983)\citenamefont {Ewen},
  \citenamefont {Taylor},\ and\ \citenamefont {Paul}}]{ewen_raman_1983}%
  \BibitemOpen
  \bibfield  {author} {\bibinfo {author} {\bibfnamefont {P.~J.~S.}\
  \bibnamefont {Ewen}}, \bibinfo {author} {\bibfnamefont {W.}~\bibnamefont
  {Taylor}},\ and\ \bibinfo {author} {\bibfnamefont {G.~L.}\ \bibnamefont
  {Paul}},\ }\bibfield  {title} {\bibinfo {title} {{A Raman scattering study of
  phase transitions in proustite (Ag$_3$AsS$_3$) and pyrargyrite
  (Ag$_3$SbS$_3$)}},\ }\href {https://doi.org/10.1088/0022-3719/16/33/019}
  {\bibfield  {journal} {\bibinfo  {journal} {J. Phys. C: Solid State Phys.}\
  }\textbf {\bibinfo {volume} {16}},\ \bibinfo {pages} {6475} (\bibinfo {year}
  {1983})}\BibitemShut {NoStop}%
\bibitem [{\citenamefont {Kihara}\ and\ \citenamefont
  {Matsumoto}(1986)}]{kihara_refinements_1986}%
  \BibitemOpen
  \bibfield  {author} {\bibinfo {author} {\bibfnamefont {K.}~\bibnamefont
  {Kihara}}\ and\ \bibinfo {author} {\bibfnamefont {T.}~\bibnamefont
  {Matsumoto}},\ }\bibfield  {title} {\bibinfo {title} {{Refinements of
  Ag$_{3}$AsSe$_3$ based on high-order thermal-motion tensors}},\ }\href
  {https://doi.org/10.1524/zkri.1986.177.14.211} {\bibfield  {journal}
  {\bibinfo  {journal} {Z. Kristallogr. Cryst. Mater}\ }\textbf {\bibinfo
  {volume} {177}},\ \bibinfo {pages} {211} (\bibinfo {year}
  {1986})}\BibitemShut {NoStop}%
\bibitem [{Sup()}]{Supp}%
  \BibitemOpen
  \href@noop {} {}\bibinfo {note} {See Supplemental Material at [URL will be
  inserted by publisher] for additional structural details, electronic band
  gaps, spin textures, and spin lifetime calculations.}\BibitemShut {Stop}%
\bibitem [{\citenamefont {Koyama}\ and\ \citenamefont
  {Rondinelli}(2022)}]{structures}%
  \BibitemOpen
  \bibfield  {author} {\bibinfo {author} {\bibfnamefont {S.}~\bibnamefont
  {Koyama}}\ and\ \bibinfo {author} {\bibfnamefont {J.~M.}\ \bibnamefont
  {Rondinelli}},\ }\href {https://doi.org/10.5281/zenodo.7039025} {\bibinfo
  {title} {{Proustite PST Data}}} (\bibinfo {year} {2022})\BibitemShut
  {NoStop}%
\bibitem [{\citenamefont {Rud’}\ \emph {et~al.}(2010)\citenamefont {Rud’},
  \citenamefont {Rud’},\ and\ \citenamefont
  {Terukov}}]{rud_development_2010}%
  \BibitemOpen
  \bibfield  {author} {\bibinfo {author} {\bibfnamefont {V.~Y.}\ \bibnamefont
  {Rud’}}, \bibinfo {author} {\bibfnamefont {Y.~V.}\ \bibnamefont {Rud’}},\
  and\ \bibinfo {author} {\bibfnamefont {E.~I.}\ \bibnamefont {Terukov}},\
  }\bibfield  {title} {\bibinfo {title} {{Development and photoelectric
  properties of In/\emph{p}-Ag$_3$AsS$_3$ surface-barrier structures}},\ }\href
  {https://doi.org/10.1134/S1063782610080129} {\bibfield  {journal} {\bibinfo
  {journal} {Semiconductors}\ }\textbf {\bibinfo {volume} {44}},\ \bibinfo
  {pages} {1025} (\bibinfo {year} {2010})}\BibitemShut {NoStop}%
\bibitem [{Note1()}]{Note1}%
  \BibitemOpen
  \bibinfo {note} {Density functional theory (DFT) calculations were performed
  using the Vienna \protect \textit {ab-initio} simulation package (VASP) \cite
  {kresse_ab_1993, kresse_efficient_1996, kresse_efficiency_1996} with a plane
  wave cutoff of \SI {350}{\electronvolt } and projector-augmented wave (PAW)
  pseudopotential\cite {kresse_ultrasoft_1999, blochl_projector_1994} with Ag
  $5s$ and $4d$; As $4s$ and $4p$; Sb $5s$ and $5p$; S $3s$ and $3p$; and Se
  $4s$ and $4p$ electrons as valence states. We utilized the PBEsol
  exchange-correlation functional with spin-orbit coupling included, unless
  specified otherwise \cite {perdew_generalized_1996, perdew_restoring_2008}.
  The Brillouin zone is sampled with a $4\times 4\times 4$ $k$-point mesh and
  integrations performed with the tetrahedon method. Structures were relaxed
  until forces were below \SI {1e-4}{\electronvolt \per \angstrom }. Electric
  polarizations were calculated using the Berry phase method \cite
  {king-smith_theory_1993}. The HSE06 hybrid functional \cite
  {krukau_influence_2006} was used for accurate bandgap calculations and dense
  $k$-point meshes were constructed for non-self-consistent field band
  dispersion and spin texture calculations. The Atomic Simulation Environment
  (ASE) was used to aid calculations and post-processing \cite
  {hjorth_larsen_atomic_2017}; LOBSTER for density of states calculations \cite
  {dronskowski_crystal_1993, deringer_crystal_2011, maintz_analytic_2013,
  maintz_lobster_2016}; and VESTA \cite {momma_vesta_2011} for structure
  visualization.}\BibitemShut {Stop}%
\bibitem [{\citenamefont {Zhang}\ \emph {et~al.}(2022)\citenamefont {Zhang},
  \citenamefont {Jiang}, \citenamefont {Multunas}, \citenamefont {Ming},
  \citenamefont {Chen}, \citenamefont {Hu}, \citenamefont {Lu}, \citenamefont
  {Pendse}, \citenamefont {Jia}, \citenamefont {Chandra}, \citenamefont {Sun},
  \citenamefont {Lu}, \citenamefont {Ping}, \citenamefont {Sundararaman},\ and\
  \citenamefont {Shi}}]{zhang_room-temperature_2022}%
  \BibitemOpen
  \bibfield  {author} {\bibinfo {author} {\bibfnamefont {L.}~\bibnamefont
  {Zhang}}, \bibinfo {author} {\bibfnamefont {J.}~\bibnamefont {Jiang}},
  \bibinfo {author} {\bibfnamefont {C.}~\bibnamefont {Multunas}}, \bibinfo
  {author} {\bibfnamefont {C.}~\bibnamefont {Ming}}, \bibinfo {author}
  {\bibfnamefont {Z.}~\bibnamefont {Chen}}, \bibinfo {author} {\bibfnamefont
  {Y.}~\bibnamefont {Hu}}, \bibinfo {author} {\bibfnamefont {Z.}~\bibnamefont
  {Lu}}, \bibinfo {author} {\bibfnamefont {S.}~\bibnamefont {Pendse}}, \bibinfo
  {author} {\bibfnamefont {R.}~\bibnamefont {Jia}}, \bibinfo {author}
  {\bibfnamefont {M.}~\bibnamefont {Chandra}}, \bibinfo {author} {\bibfnamefont
  {Y.-Y.}\ \bibnamefont {Sun}}, \bibinfo {author} {\bibfnamefont {T.-M.}\
  \bibnamefont {Lu}}, \bibinfo {author} {\bibfnamefont {Y.}~\bibnamefont
  {Ping}}, \bibinfo {author} {\bibfnamefont {R.}~\bibnamefont {Sundararaman}},\
  and\ \bibinfo {author} {\bibfnamefont {J.}~\bibnamefont {Shi}},\ }\bibfield
  {title} {\bibinfo {title} {Room-temperature electrically switchable
  spin–valley coupling in a van der waals ferroelectric halide perovskite
  with persistent spin helix},\ }\href
  {https://doi.org/10.1038/s41566-022-01016-9} {\bibfield  {journal} {\bibinfo
  {journal} {Nat. Photon.}\ }\textbf {\bibinfo {volume} {16}},\ \bibinfo
  {pages} {529} (\bibinfo {year} {2022})}\BibitemShut {NoStop}%
\bibitem [{\citenamefont {Liu}\ and\ \citenamefont
  {Sinova}(2012)}]{liu_unified_2012}%
  \BibitemOpen
  \bibfield  {author} {\bibinfo {author} {\bibfnamefont {X.}~\bibnamefont
  {Liu}}\ and\ \bibinfo {author} {\bibfnamefont {J.}~\bibnamefont {Sinova}},\
  }\bibfield  {title} {\bibinfo {title} {Unified theory of spin dynamics in a
  two-dimensional electron gas with arbitrary spin-orbit coupling strength at
  finite temperature},\ }\href {https://doi.org/10.1103/PhysRevB.86.174301}
  {\bibfield  {journal} {\bibinfo  {journal} {Phys. Rev. B}\ }\textbf {\bibinfo
  {volume} {86}},\ \bibinfo {pages} {174301} (\bibinfo {year}
  {2012})}\BibitemShut {NoStop}%
\bibitem [{\citenamefont {Shen}\ \emph {et~al.}(2022)\citenamefont {Shen},
  \citenamefont {Lin}, \citenamefont {Su}, \citenamefont {{McGahan}},
  \citenamefont {Lu}, \citenamefont {Ji}, \citenamefont {Wang}, \citenamefont
  {Wang}, \citenamefont {Mao}, \citenamefont {Guo}, \citenamefont {Park},
  \citenamefont {Wang}, \citenamefont {Tisdale}, \citenamefont {Li},
  \citenamefont {Ling}, \citenamefont {Aidala}, \citenamefont {Palacios},\ and\
  \citenamefont {Kong}}]{shen_healing_2022}%
  \BibitemOpen
  \bibfield  {author} {\bibinfo {author} {\bibfnamefont {P.-C.}\ \bibnamefont
  {Shen}}, \bibinfo {author} {\bibfnamefont {Y.}~\bibnamefont {Lin}}, \bibinfo
  {author} {\bibfnamefont {C.}~\bibnamefont {Su}}, \bibinfo {author}
  {\bibfnamefont {C.}~\bibnamefont {{McGahan}}}, \bibinfo {author}
  {\bibfnamefont {A.-Y.}\ \bibnamefont {Lu}}, \bibinfo {author} {\bibfnamefont
  {X.}~\bibnamefont {Ji}}, \bibinfo {author} {\bibfnamefont {X.}~\bibnamefont
  {Wang}}, \bibinfo {author} {\bibfnamefont {H.}~\bibnamefont {Wang}}, \bibinfo
  {author} {\bibfnamefont {N.}~\bibnamefont {Mao}}, \bibinfo {author}
  {\bibfnamefont {Y.}~\bibnamefont {Guo}}, \bibinfo {author} {\bibfnamefont
  {J.-H.}\ \bibnamefont {Park}}, \bibinfo {author} {\bibfnamefont
  {Y.}~\bibnamefont {Wang}}, \bibinfo {author} {\bibfnamefont {W.}~\bibnamefont
  {Tisdale}}, \bibinfo {author} {\bibfnamefont {J.}~\bibnamefont {Li}},
  \bibinfo {author} {\bibfnamefont {X.}~\bibnamefont {Ling}}, \bibinfo {author}
  {\bibfnamefont {K.~E.}\ \bibnamefont {Aidala}}, \bibinfo {author}
  {\bibfnamefont {T.}~\bibnamefont {Palacios}},\ and\ \bibinfo {author}
  {\bibfnamefont {J.}~\bibnamefont {Kong}},\ }\bibfield  {title} {\bibinfo
  {title} {Healing of donor defect states in monolayer molybdenum disulfide
  using oxygen-incorporated chemical vapour deposition},\ }\href
  {https://doi.org/10.1038/s41928-021-00685-8} {\bibfield  {journal} {\bibinfo
  {journal} {Nat. Electron.}\ }\textbf {\bibinfo {volume} {5}},\ \bibinfo
  {pages} {28} (\bibinfo {year} {2022})}\BibitemShut {NoStop}%
\bibitem [{\citenamefont {Acosta}\ \emph {et~al.}(2020)\citenamefont {Acosta},
  \citenamefont {Ogoshi}, \citenamefont {Fazzio}, \citenamefont {Dalpian},\
  and\ \citenamefont {Zunger}}]{Mera_Acosta_2020}%
  \BibitemOpen
  \bibfield  {author} {\bibinfo {author} {\bibfnamefont {C.~M.}\ \bibnamefont
  {Acosta}}, \bibinfo {author} {\bibfnamefont {E.}~\bibnamefont {Ogoshi}},
  \bibinfo {author} {\bibfnamefont {A.}~\bibnamefont {Fazzio}}, \bibinfo
  {author} {\bibfnamefont {G.~M.}\ \bibnamefont {Dalpian}},\ and\ \bibinfo
  {author} {\bibfnamefont {A.}~\bibnamefont {Zunger}},\ }\bibfield  {title}
  {\bibinfo {title} {The rashba scale: Emergence of band anti-crossing as a
  design principle for materials with large rashba coefficient},\ }\href
  {https://doi.org/10.1016/j.matt.2020.05.006} {\bibfield  {journal} {\bibinfo
  {journal} {Matter}\ }\textbf {\bibinfo {volume} {3}},\ \bibinfo {pages} {145}
  (\bibinfo {year} {2020})}\BibitemShut {NoStop}%
\bibitem [{\citenamefont {Kresse}\ and\ \citenamefont
  {Hafner}(1993)}]{kresse_ab_1993}%
  \BibitemOpen
  \bibfield  {author} {\bibinfo {author} {\bibfnamefont {G.}~\bibnamefont
  {Kresse}}\ and\ \bibinfo {author} {\bibfnamefont {J.}~\bibnamefont
  {Hafner}},\ }\bibfield  {title} {\bibinfo {title} {\textit{Ab initio}
  molecular dynamics for liquid metals},\ }\href
  {https://doi.org/10.1103/PhysRevB.47.558} {\bibfield  {journal} {\bibinfo
  {journal} {Phys. Rev. B}\ }\textbf {\bibinfo {volume} {47}},\ \bibinfo
  {pages} {558} (\bibinfo {year} {1993})}\BibitemShut {NoStop}%
\bibitem [{\citenamefont {Kresse}\ and\ \citenamefont
  {Furthmüller}(1996{\natexlab{a}})}]{kresse_efficient_1996}%
  \BibitemOpen
  \bibfield  {author} {\bibinfo {author} {\bibfnamefont {G.}~\bibnamefont
  {Kresse}}\ and\ \bibinfo {author} {\bibfnamefont {J.}~\bibnamefont
  {Furthmüller}},\ }\bibfield  {title} {\bibinfo {title} {Efficient iterative
  schemes for \textit{ab initio} total-energy calculations using a plane-wave
  basis set},\ }\href {https://doi.org/10.1103/PhysRevB.54.11169} {\bibfield
  {journal} {\bibinfo  {journal} {Phys. Rev. B}\ }\textbf {\bibinfo {volume}
  {54}},\ \bibinfo {pages} {11169} (\bibinfo {year}
  {1996}{\natexlab{a}})}\BibitemShut {NoStop}%
\bibitem [{\citenamefont {Kresse}\ and\ \citenamefont
  {Furthmüller}(1996{\natexlab{b}})}]{kresse_efficiency_1996}%
  \BibitemOpen
  \bibfield  {author} {\bibinfo {author} {\bibfnamefont {G.}~\bibnamefont
  {Kresse}}\ and\ \bibinfo {author} {\bibfnamefont {J.}~\bibnamefont
  {Furthmüller}},\ }\bibfield  {title} {\bibinfo {title} {Efficiency of
  ab-initio total energy calculations for metals and semiconductors using a
  plane-wave basis set},\ }\href {https://doi.org/10.1016/0927-0256(96)00008-0}
  {\bibfield  {journal} {\bibinfo  {journal} {Computational Materials Science}\
  }\textbf {\bibinfo {volume} {6}},\ \bibinfo {pages} {15} (\bibinfo {year}
  {1996}{\natexlab{b}})}\BibitemShut {NoStop}%
\bibitem [{\citenamefont {Kresse}\ and\ \citenamefont
  {Joubert}(1999)}]{kresse_ultrasoft_1999}%
  \BibitemOpen
  \bibfield  {author} {\bibinfo {author} {\bibfnamefont {G.}~\bibnamefont
  {Kresse}}\ and\ \bibinfo {author} {\bibfnamefont {D.}~\bibnamefont
  {Joubert}},\ }\bibfield  {title} {\bibinfo {title} {From ultrasoft
  pseudopotentials to the projector augmented-wave method},\ }\href
  {https://doi.org/10.1103/PhysRevB.59.1758} {\bibfield  {journal} {\bibinfo
  {journal} {Phys. Rev. B}\ }\textbf {\bibinfo {volume} {59}},\ \bibinfo
  {pages} {1758} (\bibinfo {year} {1999})}\BibitemShut {NoStop}%
\bibitem [{\citenamefont {Blöchl}(1994)}]{blochl_projector_1994}%
  \BibitemOpen
  \bibfield  {author} {\bibinfo {author} {\bibfnamefont {P.~E.}\ \bibnamefont
  {Blöchl}},\ }\bibfield  {title} {\bibinfo {title} {Projector augmented-wave
  method},\ }\href {https://doi.org/10.1103/PhysRevB.50.17953} {\bibfield
  {journal} {\bibinfo  {journal} {Phys. Rev. B}\ }\textbf {\bibinfo {volume}
  {50}},\ \bibinfo {pages} {17953} (\bibinfo {year} {1994})}\BibitemShut
  {NoStop}%
\bibitem [{\citenamefont {Perdew}\ \emph {et~al.}(1996)\citenamefont {Perdew},
  \citenamefont {Burke},\ and\ \citenamefont
  {Ernzerhof}}]{perdew_generalized_1996}%
  \BibitemOpen
  \bibfield  {author} {\bibinfo {author} {\bibfnamefont {J.~P.}\ \bibnamefont
  {Perdew}}, \bibinfo {author} {\bibfnamefont {K.}~\bibnamefont {Burke}},\ and\
  \bibinfo {author} {\bibfnamefont {M.}~\bibnamefont {Ernzerhof}},\ }\bibfield
  {title} {\bibinfo {title} {Generalized gradient approximation made simple},\
  }\href {https://doi.org/10.1103/PhysRevLett.77.3865} {\bibfield  {journal}
  {\bibinfo  {journal} {Phys. Rev. Lett.}\ }\textbf {\bibinfo {volume} {77}},\
  \bibinfo {pages} {3865} (\bibinfo {year} {1996})}\BibitemShut {NoStop}%
\bibitem [{\citenamefont {Perdew}\ \emph {et~al.}(2008)\citenamefont {Perdew},
  \citenamefont {Ruzsinszky}, \citenamefont {Csonka}, \citenamefont {Vydrov},
  \citenamefont {Scuseria}, \citenamefont {Constantin}, \citenamefont {Zhou},\
  and\ \citenamefont {Burke}}]{perdew_restoring_2008}%
  \BibitemOpen
  \bibfield  {author} {\bibinfo {author} {\bibfnamefont {J.~P.}\ \bibnamefont
  {Perdew}}, \bibinfo {author} {\bibfnamefont {A.}~\bibnamefont {Ruzsinszky}},
  \bibinfo {author} {\bibfnamefont {G.~I.}\ \bibnamefont {Csonka}}, \bibinfo
  {author} {\bibfnamefont {O.~A.}\ \bibnamefont {Vydrov}}, \bibinfo {author}
  {\bibfnamefont {G.~E.}\ \bibnamefont {Scuseria}}, \bibinfo {author}
  {\bibfnamefont {L.~A.}\ \bibnamefont {Constantin}}, \bibinfo {author}
  {\bibfnamefont {X.}~\bibnamefont {Zhou}},\ and\ \bibinfo {author}
  {\bibfnamefont {K.}~\bibnamefont {Burke}},\ }\bibfield  {title} {\bibinfo
  {title} {Restoring the density-gradient expansion for exchange in solids and
  surfaces},\ }\href {https://doi.org/10.1103/PhysRevLett.100.136406}
  {\bibfield  {journal} {\bibinfo  {journal} {Phys. Rev. Lett.}\ }\textbf
  {\bibinfo {volume} {100}},\ \bibinfo {pages} {136406} (\bibinfo {year}
  {2008})}\BibitemShut {NoStop}%
\bibitem [{\citenamefont {King-Smith}\ and\ \citenamefont
  {Vanderbilt}(1993)}]{king-smith_theory_1993}%
  \BibitemOpen
  \bibfield  {author} {\bibinfo {author} {\bibfnamefont {R.~D.}\ \bibnamefont
  {King-Smith}}\ and\ \bibinfo {author} {\bibfnamefont {D.}~\bibnamefont
  {Vanderbilt}},\ }\bibfield  {title} {\bibinfo {title} {Theory of polarization
  of crystalline solids},\ }\href {https://doi.org/10.1103/PhysRevB.47.1651}
  {\bibfield  {journal} {\bibinfo  {journal} {Phys. Rev. B}\ }\textbf {\bibinfo
  {volume} {47}},\ \bibinfo {pages} {1651} (\bibinfo {year}
  {1993})}\BibitemShut {NoStop}%
\bibitem [{\citenamefont {Krukau}\ \emph {et~al.}(2006)\citenamefont {Krukau},
  \citenamefont {Vydrov}, \citenamefont {Izmaylov},\ and\ \citenamefont
  {Scuseria}}]{krukau_influence_2006}%
  \BibitemOpen
  \bibfield  {author} {\bibinfo {author} {\bibfnamefont {A.~V.}\ \bibnamefont
  {Krukau}}, \bibinfo {author} {\bibfnamefont {O.~A.}\ \bibnamefont {Vydrov}},
  \bibinfo {author} {\bibfnamefont {A.~F.}\ \bibnamefont {Izmaylov}},\ and\
  \bibinfo {author} {\bibfnamefont {G.~E.}\ \bibnamefont {Scuseria}},\
  }\bibfield  {title} {\bibinfo {title} {Influence of the exchange screening
  parameter on the performance of screened hybrid functionals},\ }\href
  {https://doi.org/10.1063/1.2404663} {\bibfield  {journal} {\bibinfo
  {journal} {The Journal of Chemical Physics}\ }\textbf {\bibinfo {volume}
  {125}},\ \bibinfo {pages} {224106} (\bibinfo {year} {2006})}\BibitemShut
  {NoStop}%
\bibitem [{\citenamefont {Hjorth~Larsen}\ \emph {et~al.}(2017)\citenamefont
  {Hjorth~Larsen}, \citenamefont {Jørgen~Mortensen}, \citenamefont
  {Blomqvist}, \citenamefont {Castelli}, \citenamefont {Christensen},
  \citenamefont {Dułak}, \citenamefont {Friis}, \citenamefont {Groves},
  \citenamefont {Hammer}, \citenamefont {Hargus}, \citenamefont {Hermes},
  \citenamefont {Jennings}, \citenamefont {Bjerre~Jensen}, \citenamefont
  {Kermode}, \citenamefont {Kitchin}, \citenamefont {Leonhard~Kolsbjerg},
  \citenamefont {Kubal}, \citenamefont {Kaasbjerg}, \citenamefont {Lysgaard},
  \citenamefont {Bergmann~Maronsson}, \citenamefont {Maxson}, \citenamefont
  {Olsen}, \citenamefont {Pastewka}, \citenamefont {Peterson}, \citenamefont
  {Rostgaard}, \citenamefont {Schiøtz}, \citenamefont {Schütt}, \citenamefont
  {Strange}, \citenamefont {Thygesen}, \citenamefont {Vegge}, \citenamefont
  {Vilhelmsen}, \citenamefont {Walter}, \citenamefont {Zeng},\ and\
  \citenamefont {Jacobsen}}]{hjorth_larsen_atomic_2017}%
  \BibitemOpen
  \bibfield  {author} {\bibinfo {author} {\bibfnamefont {A.}~\bibnamefont
  {Hjorth~Larsen}}, \bibinfo {author} {\bibfnamefont {J.}~\bibnamefont
  {Jørgen~Mortensen}}, \bibinfo {author} {\bibfnamefont {J.}~\bibnamefont
  {Blomqvist}}, \bibinfo {author} {\bibfnamefont {I.~E.}\ \bibnamefont
  {Castelli}}, \bibinfo {author} {\bibfnamefont {R.}~\bibnamefont
  {Christensen}}, \bibinfo {author} {\bibfnamefont {M.}~\bibnamefont {Dułak}},
  \bibinfo {author} {\bibfnamefont {J.}~\bibnamefont {Friis}}, \bibinfo
  {author} {\bibfnamefont {M.~N.}\ \bibnamefont {Groves}}, \bibinfo {author}
  {\bibfnamefont {B.}~\bibnamefont {Hammer}}, \bibinfo {author} {\bibfnamefont
  {C.}~\bibnamefont {Hargus}}, \bibinfo {author} {\bibfnamefont {E.~D.}\
  \bibnamefont {Hermes}}, \bibinfo {author} {\bibfnamefont {P.~C.}\
  \bibnamefont {Jennings}}, \bibinfo {author} {\bibfnamefont {P.}~\bibnamefont
  {Bjerre~Jensen}}, \bibinfo {author} {\bibfnamefont {J.}~\bibnamefont
  {Kermode}}, \bibinfo {author} {\bibfnamefont {J.~R.}\ \bibnamefont
  {Kitchin}}, \bibinfo {author} {\bibfnamefont {E.}~\bibnamefont
  {Leonhard~Kolsbjerg}}, \bibinfo {author} {\bibfnamefont {J.}~\bibnamefont
  {Kubal}}, \bibinfo {author} {\bibfnamefont {K.}~\bibnamefont {Kaasbjerg}},
  \bibinfo {author} {\bibfnamefont {S.}~\bibnamefont {Lysgaard}}, \bibinfo
  {author} {\bibfnamefont {J.}~\bibnamefont {Bergmann~Maronsson}}, \bibinfo
  {author} {\bibfnamefont {T.}~\bibnamefont {Maxson}}, \bibinfo {author}
  {\bibfnamefont {T.}~\bibnamefont {Olsen}}, \bibinfo {author} {\bibfnamefont
  {L.}~\bibnamefont {Pastewka}}, \bibinfo {author} {\bibfnamefont
  {A.}~\bibnamefont {Peterson}}, \bibinfo {author} {\bibfnamefont
  {C.}~\bibnamefont {Rostgaard}}, \bibinfo {author} {\bibfnamefont
  {J.}~\bibnamefont {Schiøtz}}, \bibinfo {author} {\bibfnamefont
  {O.}~\bibnamefont {Schütt}}, \bibinfo {author} {\bibfnamefont
  {M.}~\bibnamefont {Strange}}, \bibinfo {author} {\bibfnamefont {K.~S.}\
  \bibnamefont {Thygesen}}, \bibinfo {author} {\bibfnamefont {T.}~\bibnamefont
  {Vegge}}, \bibinfo {author} {\bibfnamefont {L.}~\bibnamefont {Vilhelmsen}},
  \bibinfo {author} {\bibfnamefont {M.}~\bibnamefont {Walter}}, \bibinfo
  {author} {\bibfnamefont {Z.}~\bibnamefont {Zeng}},\ and\ \bibinfo {author}
  {\bibfnamefont {K.~W.}\ \bibnamefont {Jacobsen}},\ }\bibfield  {title}
  {\bibinfo {title} {The atomic simulation environment—a python library for
  working with atoms},\ }\href {https://doi.org/10.1088/1361-648X/aa680e}
  {\bibfield  {journal} {\bibinfo  {journal} {J. Phys.: Condens. Matter}\
  }\textbf {\bibinfo {volume} {29}},\ \bibinfo {pages} {273002} (\bibinfo
  {year} {2017})}\BibitemShut {NoStop}%
\bibitem [{\citenamefont {Dronskowski}\ and\ \citenamefont
  {Bloechl}(1993)}]{dronskowski_crystal_1993}%
  \BibitemOpen
  \bibfield  {author} {\bibinfo {author} {\bibfnamefont {R.}~\bibnamefont
  {Dronskowski}}\ and\ \bibinfo {author} {\bibfnamefont {P.~E.}\ \bibnamefont
  {Bloechl}},\ }\bibfield  {title} {\bibinfo {title} {Crystal orbital hamilton
  populations ({COHP}): energy-resolved visualization of chemical bonding in
  solids based on density-functional calculations},\ }\href
  {https://doi.org/10.1021/j100135a014} {\bibfield  {journal} {\bibinfo
  {journal} {J. Phys. Chem.}\ }\textbf {\bibinfo {volume} {97}},\ \bibinfo
  {pages} {8617} (\bibinfo {year} {1993})}\BibitemShut {NoStop}%
\bibitem [{\citenamefont {Deringer}\ \emph {et~al.}(2011)\citenamefont
  {Deringer}, \citenamefont {Tchougréeff},\ and\ \citenamefont
  {Dronskowski}}]{deringer_crystal_2011}%
  \BibitemOpen
  \bibfield  {author} {\bibinfo {author} {\bibfnamefont {V.~L.}\ \bibnamefont
  {Deringer}}, \bibinfo {author} {\bibfnamefont {A.~L.}\ \bibnamefont
  {Tchougréeff}},\ and\ \bibinfo {author} {\bibfnamefont {R.}~\bibnamefont
  {Dronskowski}},\ }\bibfield  {title} {\bibinfo {title} {Crystal orbital
  hamilton population ({COHP}) analysis as projected from plane-wave basis
  sets},\ }\href {https://doi.org/10.1021/jp202489s} {\bibfield  {journal}
  {\bibinfo  {journal} {J. Phys. Chem. A}\ }\textbf {\bibinfo {volume} {115}},\
  \bibinfo {pages} {5461} (\bibinfo {year} {2011})}\BibitemShut {NoStop}%
\bibitem [{\citenamefont {Maintz}\ \emph {et~al.}(2013)\citenamefont {Maintz},
  \citenamefont {Deringer}, \citenamefont {Tchougréeff},\ and\ \citenamefont
  {Dronskowski}}]{maintz_analytic_2013}%
  \BibitemOpen
  \bibfield  {author} {\bibinfo {author} {\bibfnamefont {S.}~\bibnamefont
  {Maintz}}, \bibinfo {author} {\bibfnamefont {V.~L.}\ \bibnamefont
  {Deringer}}, \bibinfo {author} {\bibfnamefont {A.~L.}\ \bibnamefont
  {Tchougréeff}},\ and\ \bibinfo {author} {\bibfnamefont {R.}~\bibnamefont
  {Dronskowski}},\ }\bibfield  {title} {\bibinfo {title} {Analytic projection
  from plane-wave and {PAW} wavefunctions and application to chemical-bonding
  analysis in solids},\ }\href {https://doi.org/10.1002/jcc.23424} {\bibfield
  {journal} {\bibinfo  {journal} {J. Comput. Chem.}\ }\textbf {\bibinfo
  {volume} {34}},\ \bibinfo {pages} {2557} (\bibinfo {year}
  {2013})}\BibitemShut {NoStop}%
\bibitem [{\citenamefont {Maintz}\ \emph {et~al.}(2016)\citenamefont {Maintz},
  \citenamefont {Deringer}, \citenamefont {Tchougréeff},\ and\ \citenamefont
  {Dronskowski}}]{maintz_lobster_2016}%
  \BibitemOpen
  \bibfield  {author} {\bibinfo {author} {\bibfnamefont {S.}~\bibnamefont
  {Maintz}}, \bibinfo {author} {\bibfnamefont {V.~L.}\ \bibnamefont
  {Deringer}}, \bibinfo {author} {\bibfnamefont {A.~L.}\ \bibnamefont
  {Tchougréeff}},\ and\ \bibinfo {author} {\bibfnamefont {R.}~\bibnamefont
  {Dronskowski}},\ }\bibfield  {title} {\bibinfo {title} {{LOBSTER}: A tool to
  extract chemical bonding from plane-wave based {DFT}: Tool to extract
  chemical bonding},\ }\href {https://doi.org/10.1002/jcc.24300} {\bibfield
  {journal} {\bibinfo  {journal} {J. Comput. Chem.}\ }\textbf {\bibinfo
  {volume} {37}},\ \bibinfo {pages} {1030} (\bibinfo {year}
  {2016})}\BibitemShut {NoStop}%
\bibitem [{\citenamefont {Momma}\ and\ \citenamefont
  {Izumi}(2011)}]{momma_vesta_2011}%
  \BibitemOpen
  \bibfield  {author} {\bibinfo {author} {\bibfnamefont {K.}~\bibnamefont
  {Momma}}\ and\ \bibinfo {author} {\bibfnamefont {F.}~\bibnamefont {Izumi}},\
  }\bibfield  {title} {\bibinfo {title} {\textit{{VESTA} 3} for
  three-dimensional visualization of crystal, volumetric and morphology data},\
  }\href {https://doi.org/10.1107/S0021889811038970} {\bibfield  {journal}
  {\bibinfo  {journal} {J Appl Crystallogr}\ }\textbf {\bibinfo {volume}
  {44}},\ \bibinfo {pages} {1272} (\bibinfo {year} {2011})}\BibitemShut
  {NoStop}%
\end{thebibliography}%
\end{document}